\documentclass[epj]{svjour}
\newcommand{\beq}{\begin{equation}} 
\newcommand{\eeq}{\end{equation}} 
\newcommand{\beqa}{\begin{eqnarray}} 
\newcommand{\eeqa}{\end{eqnarray}} 
\newcommand{\beqan}{\begin{eqnarray*}} 
\newcommand{\eeqan}{\end{eqnarray*}} 
\newcommand{\ba}{\begin{array}} 
\newcommand{\ea}{\end{array}} 
\newcommand{\no}{\nonumber}

\newcommand{\ol}{\overline}

\newcommand{\ve}{\varepsilon}

\newcommand{\dg}{\dagger} 
\newcommand{\wt}{\widetilde} 
\newcommand{\wh}{\widehat}

\newcommand{\D}{{\cal D}}

\newcommand{\cL}{{\cal L}}

\newcommand{\cO}{{\cal O}} 
 
\newcommand{\Q}{{\cal Q}}

\newcommand{\nn}{\nonumber \\}

\newcommand{\bea}{\begin{eqnarray}} 
\newcommand{\eea}{\end{eqnarray}}

\begin{document}

\onecolumn

\thispagestyle{empty}

\begin{flushright} 
UWThPh-2001-44\\ 
IFIC/01-55 \\ 
CPT-2001/P.4248\\
\end{flushright} 
\vspace{2.5cm} 

\begin{center}
{\huge {\bf 
Radiative corrections to $\mbox{\boldmath $K_{\ell 3}$}$ decays $^*$ }}
\end{center}   
\vskip 2.0cm

\begin{center}
{\large V. Cirigliano$^{1,2}$, M. Knecht$^3$, H. Neufeld$^1$, 
H. Rupertsberger$^1$, P. Talavera$^3$ } \\

\vskip 0.5cm

$^1$ Institut f\"ur Theoretische Physik, Universit\"at 
Wien, Boltzmanngasse 5, A-1090 Wien, Austria \\
$^2$ Departament de F\'{\i}sica Te\`{o}rica, IFIC, Universitat de 
Val\`{e}ncia -- CSIC, \\
Edifici d'Instituts de Paterna,  
Apt. Correus 22085, E-46071 Val\`{e}ncia, Spain \\
$^3$ Centre de Physique Th\'eorique, CNRS 
Luminy, Case 907, F-13288 Marseille - Cedex 9, France \\
\end{center}   
\vskip 1.5cm

\begin{center}
{\bf Abstract}
\end{center}
\noindent
We present a complete calculation of the $K_{\ell 3}$ decays 
$K^+ \to \pi^0 \ell^+ \nu_{\ell}$ and $K^0 \to \pi^- \ell^+ \nu_\ell$ to
$\cO(p^4, (m_d-m_u) p^2, e^2 p^2)$ in chiral perturbation theory with 
virtual photons and leptons. We introduce the concept of generalized form 
factors and kinematical densities in the presence of 
electromagnetism, and propose a possible treatment of the real photon 
emission in $K^+_{\ell 3}$ decays. We illustrate our results by applying 
them to the extraction of the Kobayashi--Maskawa matrix element $|V_{us}|$ 
from the experimental $K^+_{e3}$ decay parameters.
\vskip 0.2cm

                 weak-interaction processes - 
      12.15.Hh Determination of Kobayashi--Maskawa matrix elements - 
      12.39.Fe Chiral Lagrangians 

\vfill 
\noindent *~Work supported in part by TMR, EC-Contract  
No. ERBFMRX-CT980169 (EURODA$\Phi$NE).

\newpage

\twocolumn

\section{Introduction}
\label{sec: Introduction}
\renewcommand{\theequation}{\arabic{section}.\arabic{equation}}
\setcounter{equation}{0}

Semileptonic kaon decays have played a central role in our
understanding of flavour physics. In particular, the $K_{e3}$ decay
mode is often presented as the most accurate and
theoretically cleanest source for the extraction of the Kobayashi--Maskawa 
matrix element
$|V_{us}|$ \cite{PDG}.  However, as new high precision experiments are
planned and the theoretical tool of effective field theory has been
pushed to higher orders, an update of the theoretical analysis of such
decays is needed. In particular, it is interesting to improve as much
as possible the theoretical analysis underlying the extraction of
$|V_{us}|$.  This is presently based on the following ingredients: a
calculation of the hadronic form factors \cite{gl852}
at order $p^4$ in chiral perturbation theory (CHPT) \cite{chpt,GL85}, 
supplemented by a model-dependent estimate of the order
$p^6$ effects \cite{lr84}.  As for the radiative corrections, the
model independent short distance leading logarithms \cite{Sirlin,MS93} 
have
been included, together with a model dependent estimate of the long
distance contributions \cite{gin66,gin67,gin68,gin70}.

It is the purpose of this paper to work out the radiative corrections
to the $K_{\ell 3}$ decays ($\ell=e,\mu$) within the framework of CHPT, 
the effective theory of the standard model at low energy.  The appropriate 
formalism for including
virtual photons in purely mesonic low-energy processes  has been
presented in \cite{urech,nr95,nr96}. This scheme was then extended for the 
treatment of semileptonic interactions by the additional inclusion of 
virtual leptons \cite{lept}.
The goal of our analysis is to obtain the size of such corrections 
and quantify the theoretical uncertainty to be assigned to them. 
This will allow us to understand how well one can know $|V_{us}|$ 
once new high statistics experiments collect data and the 
full order $p^6$ CHPT analysis becomes available \cite{bi01}. 

The outline of the paper is as follows. In Sect. \ref{sec: SM} we 
recapitulate CHPT in the presence of virtual photons and leptons. The 
basic $K_{\ell 3}$ phenomenology in the absence of electromagnetism is 
reviewed in Sect. \ref{sec: Kl3-phen}. The structure of the radiative 
corrections to $K_{\ell 3}$ decays is discussed in Sect. \ref{sec: 
structure}. The $K_{\ell 3}$ amplitudes to $\cO(p^4)$ and $\cO(e^2 p^2)$ 
are calculated in Sect. \ref{sec: CHPT}. In Sect. \ref{sec: corr} we 
propose a specific treatment of the real photon emission in the $K^+_{\ell 
3}$ case. In Sect. \ref{sec: applic} we illustrate our general considerations
by a numerical study of the $K_{e 3}^+$ decay and the description of 
a procedure to extract the 
Kobayashi--Maskawa matrix element $|V_{us}|$ from experimental data. Our 
conclusions are summarized in Sect. \ref{sec: Conclusions}. Appendix 
\ref{appA} contains a summary of integrals appearing in various mesonic 
one-loop amplitudes and Appendix \ref{appB} collects several photonic loop 
functions. In Appendix \ref{appC} we report a set of coefficients 
appearing in the next-to-leading order expansion of the form factors.

\section{The standard model at low energies}
\label{sec: SM}
\renewcommand{\theequation}{\arabic{section}.\arabic{equation}}
\setcounter{equation}{0}

The appropriate theoretical framework for the analysis of electromagnetic 
effects in semileptonic kaon decays is a low-energy effective 
quantum field theory where the  asymptotic states consist of the 
pseudoscalar octet, the photon and the light leptons \cite{lept}. 
The corresponding lowest-order effective Lagrangian is given by
\beqa \label{Leff}
\cL_{\rm eff} &=& \frac{F_0^2}{4} \; \langle u_\mu u^\mu + \chi_+\rangle +
e^2 F_0^4 Z \langle \Q_{\rm L}^{\rm em} \Q_{\rm R}^{\rm em}\rangle 
- \frac{1}{4} F_{\mu\nu} F^{\mu\nu}                \no \\
&& \mbox{} + \sum_\ell
[ \bar \ell (i \! \not\!\partial + e \! \not\!\!A - m_\ell)\ell +
\ol{\nu_{\ell \rm L}} \, i \! \not\!\partial \nu_{\ell \rm L}]  .
\eeqa
The symbol $\langle \; \rangle$ denotes the trace in three-dimensional 
flavour space, and
\beq \label{umu}
u_\mu = i [u^\dg (\partial_\mu - i r_\mu)u - u
(\partial_\mu - i l_\mu)u^\dg]  .
\eeq
The photon field $A_\mu$ and the leptons $\ell,\nu_\ell$ ($\ell = e,\mu$)
are contained in (\ref{umu})
by adding appropriate terms to the usual external vector and axial-vector
sources $v_\mu$, $a_\mu$:
\beqa \label{sources}
l_\mu &=& v_\mu - a_\mu - e Q_{\rm L}^{\rm em} A_\mu + \sum_\ell
(\bar \ell \gamma_\mu \nu_{\ell \rm L} Q_{\rm L}^{\rm w} + \ol{\nu_{\ell 
\rm L}} 
\gamma_\mu \ell
Q_{\rm L}^{{\rm w}\dg})  , \no \\
r_\mu &=& v_\mu + a_\mu - e Q_{\rm R}^{\rm em} A_\mu  .
\eeqa
The $3 \times 3$ matrices $Q_{\rm L,R}^{\rm em}$, $Q_{\rm L}^{\rm w}$ are 
spurion 
fields.
At the end, one identifies 
$Q_{\rm L,R}^{\rm em}$ with the quark charge matrix
\beq \label{Qem}
Q^{\rm em} = \left[ \ba{ccc} 2/3 & 0 & 0 \\ 0 & -1/3 & 0 \\ 0 & 0 & -1/3 \ea
\right],
\eeq
whereas the weak spurion is taken at
\beq \label{Qw}
Q_{\rm L}^{\rm w} = - 2 \sqrt{2}\; G_{\rm F} \left[ \ba{ccc}
0 & V_{ud} & V_{us} \\ 0 & 0 & 0 \\ 0 & 0 & 0 \ea \right],
\eeq
where $G_{\rm F}$ is the Fermi coupling constant and $V_{ud}$, $V_{us}$ 
are
Kobayashi--Maskawa matrix elements.
For the construction of the effective Lagrangian it is also convenient
to define
\beq
\Q_{\rm L}^{\rm em,w} := u Q_{\rm L}^{\rm em,w} u^\dg  , \qquad
\Q_{\rm R}^{\rm em} := u^\dg Q_{\rm R}^{\rm em} u  .
\eeq
$F_0$ denotes the pion decay constant in the chiral limit and in the 
absence of electroweak interactions.
Explicit chiral symmetry breaking is included in
$\chi_+ = u^\dg \chi u^\dg + u \chi^\dg u$ where $\chi$ is 
proportional to the quark mass matrix: 
\beq \label{chi}
\chi = 2 B_0 \left[ \ba{ccc} m_u & 0 & 0 \\ 0 & m_d & 0 \\ 0 & 0 & m_s 
\ea
\right].
\eeq
The factor $B_0$ is related to the quark condensate in the chiral limit by 
$\langle 0 | \ol q q | 0 \rangle = -F_0^2 B_0$. This leads to the
(lowest-order) expressions for the pseudoscalar masses
\beqa
 M^2_{\pi^\pm} &=& 2B_0 \wh m + 2 e^2 Z F_0^2  , \no \\
 M^2_{\pi^0} &=&  2B_0 \wh m  , \no \\
 M^2_{K^\pm} &=& B_0 \left[ (m_s + \wh m) - \frac{2\ve^{(2)}}{\sqrt{3}}
(m_s - \wh m)\right] + 2e^2 Z F_0^2  , \no \\
 M^2_{\stackrel{(-)}{K}{}^0} &=& B_0 \left[(m_s + \wh m) + 
\frac{2\ve^{(2)}}{\sqrt{3}} (m_s - \wh m)\right]  , \no \\
 M^2_\eta &=& \frac{4}{3} B_0 \left( m_s + \frac{\wh m}{2}\right)  .
\label{treemass}
\eeqa
The tree-level mixing angle $\ve^{(2)}$ is given by
\beq
\ve^{(2)} = \frac{\sqrt{3}}{4} \; \frac{m_d - m_u}{m_s - \wh m}  ,
\label{epsilon}
\eeq 
the symbol $\wh m$ stands for the mean value of the light quark masses,
\beq
\wh m = \frac{1}{2} (m_u + m_d)  .
\eeq
For later use, we also denote the
isospin limits ($m_u = m_d$, $e = 0$) of the pion mass and the
kaon mass, respectively, by
\beq
 M^2_{\pi} =  2B_0 \wh m  , \quad 
 M^2_K = B_0 (m_s + \wh m)  .
\eeq
Using (\ref{treemass}), the numerical value of the coupling constant $Z 
\simeq 0.8$ can be determined from the mass difference of the charged pions.

In our calculation we include terms in the low-energy expansion 
up to order $p^4, (m_u - m_d) p^2$ and $e^2 p^2$.

The most general local action at next-to-leading order can be written as 
the sum of four terms, $\cL_{p^4} + \cL_{e^2 p^2} + \cL_{\rm lept} + 
\cL_\gamma$. The first one, $\cL_{p^4}$ is the well-known Gasser-Leutwyler 
Lagrangian \cite{GL85} in the presence of the generalized external sources 
introduced in (\ref{sources}): 
\beqa
\cL_{p^4} &=& L_1 \; \langle u_\mu u^\mu \rangle^2 + L_2 \; \langle u_\mu
u^\nu\rangle \; \langle u^\mu u_\nu\rangle \no \\
&& \mbox{} + L_3 \; \langle u_\mu u^\mu u_\nu u^\nu\rangle + 
L_4 \; \langle u_\mu u^\mu\rangle \; \langle \chi_+\rangle \no \\
&& \mbox{} + L_5 \; \langle u_\mu u^\mu \chi_+\rangle + 
L_6 \; \langle \chi_+\rangle^2 + L_7 \; \langle \chi_-\rangle^2 \no \\
&& \mbox{} + \frac{1}{4} (2L_8 + L_{12}) \langle \chi_+^2\rangle + 
\frac{1}{4} (2L_8 - L_{12}) \langle \chi_-^2\rangle \no \\
&& \mbox{} - iL_9 \; \langle f_+^{\mu\nu} u_\mu u_\nu\rangle + 
\frac{1}{4} (L_{10} + 2L_{11}) \langle f_{+\mu\nu} f_+^{\mu\nu}\rangle
\no \\
&& \mbox{} - \frac{1}{4} (L_{10} - 2L_{11}) 
\langle f_{-\mu\nu} f_-^{\mu\nu}\rangle  , \label{L4}
\eeqa
with
\beqa
f_{\pm}^{\mu \nu} &=& u F_{\rm L}^{\mu \nu} u^{\dg} \pm 
                       u^{\dg} F_{\rm R}^{\mu \nu} u  , \nn
F_{\rm L}^{\mu \nu} &=& \partial^{\mu} l^{\nu} - \partial^{\nu} l^{\mu}
                  - i [l^\mu,l^\nu]  , \nn
F_{\rm R}^{\mu \nu} &=& \partial^{\mu} r^{\nu} - \partial^{\nu} r^{\mu}
                  - i [r^\mu,r^\nu]  .
\eeqa
The second term, $\cL_{e^2 p^2}$, denotes the interaction of 
dynamical photons with hadronic degrees of freedom 
\cite{urech,nr95,nr96}. Its expression is:
\beqa
\cL_{e^2p^2} &=&
e^2 F_0^2 \bigg\{ \frac{1}{2} K_1 \; \langle (\Q^{\rm em}_{\rm L})^2 +
(\Q^{\rm em}_{\rm R})^2\rangle \; \langle u_\mu  
u^\mu\rangle  \no \\
&& \mbox{} + K_2 \; \langle \Q^{\rm em}_{\rm L} \Q^{\rm em}_{\rm R}\rangle 
\; \langle u_\mu u^\mu
\rangle \no \\
&& \mbox{} - K_3 \; [\langle \Q^{\rm em}_{\rm L} u_\mu\rangle 
\; \langle \Q^{\rm em}_{\rm L} u^\mu
\rangle + \langle \Q^{\rm em}_{\rm R} u_\mu\rangle 
\; \langle \Q^{\rm em}_{\rm R} u^\mu\rangle ] \no \\
&& \mbox{} + K_4 \; \langle \Q^{\rm em}_{\rm L} u_\mu\rangle 
\; \langle \Q^{\rm em}_{\rm R} u^\mu \rangle
\no \\
&& \mbox{} + K_5 \; \langle[(\Q^{\rm em}_{\rm L})^2 + (\Q^{\rm 
em}_{\rm R})^2] 
u_\mu u^\mu\rangle \no \\
&& \mbox{} + K_6 \; \langle (\Q^{\rm em}_{\rm L} \Q^{\rm em}_{\rm R} + 
\Q^{\rm em}_{\rm R} \Q^{\rm em}_{\rm L}) u_\mu u^\mu\rangle \no \\
&& \mbox{} + \frac{1}{2} K_7 \; \langle (\Q^{\rm em}_{\rm L})^2 
+ (\Q^{\rm em}_{\rm R})^2\rangle \; \langle \chi_+\rangle
\no \\
&& \mbox{} + K_8\; \langle \Q^{\rm em}_{\rm L} \Q^{\rm em}_{\rm R}\rangle 
\; \langle \chi_+\rangle \no \\
&& \mbox{}+ K_9 \; \langle [(\Q^{\rm em}_{\rm L})^2 + (\Q^{\rm 
em}_{\rm R})^2] 
\chi_+\rangle \no \\
&& \mbox{} + K_{10}\; \langle(\Q^{\rm em}_{\rm L} \Q^{\rm em}_{\rm R} 
+ \Q^{\rm em}_{\rm R} \Q^{\rm em}_{\rm L}) \chi_+\rangle \no \\
&& \mbox{} - K_{11} \; \langle(\Q^{\rm em}_{\rm L} \Q^{\rm em}_{\rm R} 
- \Q^{\rm em}_{\rm R} \Q^{\rm em}_{\rm L}) \chi_-\rangle \no \\
&& \mbox{}- iK_{12}\; \langle[(\wh \nabla_\mu \Q^{\rm em}_{\rm L}) \Q^{\rm 
em}_{\rm L} -
\Q^{\rm em}_{\rm L} \wh \nabla_\mu \Q^{\rm em}_{\rm L} \no \\ 
&& \ \ \ \ \ \ \ \ \ \ {}- (\wh \nabla_\mu \Q^{\rm em}_{\rm R}) \Q^{\rm 
em}_{\rm 
R} + 
\Q^{\rm em}_{\rm R} \wh \nabla_\mu \Q^{\rm em}_{\rm R}] u^\mu\rangle \no 
\\
&& \mbox{}+ K_{13} \; \langle (\wh \nabla_\mu \Q^{\rm em}_{\rm L}) 
(\wh \nabla^\mu \Q^{\rm em}_{\rm R})
\rangle \no \\
&&  \mbox{} + K_{14} \; \langle (\wh \nabla_\mu \Q^{\rm em}_{\rm L}) 
(\wh \nabla^\mu \Q^{\rm em}_{\rm L}) \no \\
&& \ \ \ \ \ \ \ \ \ {}+ (\wh \nabla_\mu \Q^{\rm em}_{\rm R}) (\wh 
\nabla^\mu 
\Q^{\rm 
em}_{\rm R})\rangle  
\vphantom{\frac{1}{2} K_1}
\bigg\}  , 
\label{LE2P2}
\eeqa
where
\beqa \label{covder1}
\wh \nabla_\mu \Q^{\rm em}_{\rm L} &=& \nabla_\mu \Q^{\rm em}_{\rm L} 
+ \frac{i}{2} [u_\mu,\Q^{\rm em}_{\rm L}] \nn 
&=& u D_\mu Q_{\rm L}^{\rm em} u^\dg  , \no \\
\wh \nabla_\mu \Q^{\rm em}_{\rm R} &=& \nabla_\mu \Q^{\rm em}_{\rm R} 
- \frac{i}{2} [u_\mu,\Q^{\rm em}_{\rm R}] \nn
&=& u^\dg D_\mu Q^{\rm em}_{\rm R} u  ,
\eeqa
with
\beqa \label{covder2}
&& D_\mu Q^{\rm em}_{\rm L} = \partial_\mu Q^{\rm em}_{\rm L} 
- i[l_\mu,Q^{\rm em}_{\rm L}]  , \nn
&& D_\mu Q^{\rm em}_{\rm R} = \partial_\mu Q^{\rm em}_{\rm R} - 
i[r_\mu,Q^{\rm 
em}_{\rm R}] 
 .
\eeqa
The presence of virtual leptons requires also the addition of the 
``leptonic'' term \cite{lept}
\beqa \label{Llept}
\cL_{\rm lept} &=& 
e^2 \sum_{\ell} \Big\{ {F_0}^2 \Big[  
X_1 \ol{\ell} \gamma_\mu \nu_{\ell L} 
\langle u^\mu  \{ \Q_{\rm R}^{\rm em}, \Q_{\rm L}^{\rm w} \} \rangle 
\nn [-5pt]
&& \qquad \quad \quad \ 
{} + X_2 \ol{\ell} \gamma_\mu \nu_{\ell L} 
\langle u^\mu  [\Q_{\rm R}^{\rm em}, \Q_{\rm L}^{\rm w}] \rangle
\nn
&& \qquad \quad \quad \
{} + X_3 m_\ell \ol{\ell} \nu_{\ell L} \langle \Q_{\rm L}^{\rm w} 
\Q_{\rm R}^{\rm 
em} \rangle
\nn
&& \qquad \quad \quad \
{} + i X_4 \ol{\ell} \gamma_\mu \nu_{\ell L} 
\langle \Q_{\rm L}^{\rm w} \wh \nabla^\mu  \Q_{\rm L}^{\rm em} \rangle
\nn
&& \qquad \quad \quad \
{} + i X_5 \ol{\ell} \gamma_\mu \nu_{\ell L} 
\langle \Q_{\rm L}^{\rm w} \wh \nabla^\mu  \Q_{\rm R}^{\rm em} \rangle 
+ h.c. \Big]  \nn
&& \quad  \quad 
\ \ {} + X_6 \bar \ell (i \! \not\!\partial + e \! \not\!\!A )\ell
\nn
&& \quad  \quad
\ \ {} + X_7 m_\ell \ol \ell  \ell \Big\}  . 
\eeqa
In $\cL_{\rm lept}$  we consider only terms quadratic in the lepton
fields and at most linear in $G_{\rm F}$. 
Finally, also a photon Lagrangian
\beq
\cL_{\gamma} = e^2 X_8 F_{\mu \nu} F^{\mu \nu}  , \quad 
F_{\mu \nu} = \partial_{\mu} A_{\nu} - \partial_{\nu} A_{\mu}  ,
\eeq
has to be added \cite{lept}. 

The low-energy couplings $L_i$, $K_i$, $X_i$ arising here are
divergent (except $L_3$, $L_7$, $K_7$, $K_{13}$, $K_{14}$ and $X_1$). 
They absorb the poles of the
one-loop graphs via the renormalization
\beqa \label{renorm}
L_i &=& L_i^r(\mu) + \Gamma_i \Lambda(\mu)  , \quad i=1,\ldots,12  , \nn
K_i &=& K_i^r(\mu) + \Sigma_i \Lambda(\mu)  , \quad i=1,\ldots,14  , \nn
X_i &=& X_i^r(\mu) + \Xi_i \Lambda(\mu)  , \quad i=1,\ldots,8  ,
\eeqa
with
\beq \label{Lambda} 
\Lambda(\mu) = \frac{\mu^{d-4}}{(4\pi)^2} \left\{ \frac{1}{d-4} -
\frac{1}{2} [\ln (4\pi) + \Gamma'(1) + 1]\right\} 
\eeq
in the dimensional regularization scheme. The coefficients
$\Gamma_i$ and $\Sigma_i$ can be found in \cite{GL85} and in
\cite{urech}, respectively. Their values are not modified by the
presence of virtual leptons as long as contributions of
$\cO(G_{\rm F}^2)$ are neglected. 
The ``leptonic'' coefficients $\Xi_i$  have been 
determined rather recently in \cite{lept} by using super-heat-kernel 
techniques \cite{shk}.

In order to match our low-energy effective theory to the standard model of 
strong and electroweak interactions, we have to specify the precise 
physical meaning of the parameter $G_{\rm F}$. In the presence of 
electromagnetism, this identification is somewhat ambiguous. The shift
\beq \label{shift1}
G_{\rm F} \to G_{\rm F} ( 1 + e^2 \delta )
\eeq
induces the change (see Eq. (4.8) of \cite{lept})
\beq \label{shift2}
\frac{F_0^2}{4} \; \langle u_\mu u^\mu \rangle 
\to
\frac{F_0^2}{4} \; \langle u_\mu u^\mu \rangle 
- 2 e^2 \delta \sum_\ell
\bar \ell (i \! \not\!\partial + e \! \not\!\!A - m_\ell)\ell  ,
\eeq
corresponding to 
\beq
X_6 \to X_6 - 2 \delta  , \qquad X_7 \to X_7 + 2 \delta  .
\eeq
In other words, some part of the electromagnetic contributions  
may always be shuffled from $X_6$ to $G_{\rm F}$ or vice versa. (The 
coupling 
constant $X_7$ does not appear in observable quantities as it is always 
absorbed by the mass renormalization of the charged leptons.)

Following \cite{lept}, we identify $G_{\rm F}$ with the muon decay 
constant. 
To  order $\alpha$, $G_{\rm F}$ can be related to the 
measured muon decay width by \cite{KS59}
\beq
\Gamma(\mu\to {\mbox{all}}) = \frac{G_{\rm F}^2m_\mu^5}{192\pi^3}\,
f \! \left(\frac{m_e^2}{m_\mu^2}\right)\Big[1 + \frac{\alpha}{2\pi}
\Big( \frac{25}{4}-\pi^2 \Big) + \cO(\alpha^2) \Big]  ,
\eeq
with $f(x) = 1 -8x -12x^2\ln x +8x^3 - x^4$. 
With this choice of $G_{\rm F}$, the (universal) short-distance 
electromagnetic 
correction \cite{Sirlin,MS93} of semileptonic charged current amplitudes 
is fully contained in the coupling constant $X_6$. To display this 
dependence explicitly, we pull out the short-distance part $X_6^{\rm SD}$ 
by the decomposition
\beq \label{X6SD}
X_6^r(\mu) = X_6^{\rm SD} + \wt{X}_6^{r} (\mu)  ,
\eeq
where
\beq \label{SEW}
e^2 X_6^{\rm SD} = -\frac{e^2}{4 \pi^2} \log \frac{M_Z^2}{M_{\rho}^2}
= 1 - S_{\rm EW} (M_\rho,M_Z)  ,
\eeq
which defines \cite{MS93} also the short-distance enhancement factor 
$S_{\rm EW}(M_\rho,M_Z)$ (to leading order). With this splitting of $X_6$, 
we expect its ``long-distance part'' $\wt{X}_6^r (M_\rho)$ having the 
typical size of a low-energy 
constant in CHPT. 

\section{Basic $\mbox{\boldmath $K_{\ell 3}$}$ phenomenology}
\label{sec: Kl3-phen} 
\renewcommand{\theequation}{\arabic{section}.\arabic{equation}}
\setcounter{equation}{0}

In this section we report the basic formulae of $K_{\ell 3}$ 
phenomenology in the absence of radiative corrections 
\footnote{In the present section $M_K$ and $M_\pi$ stand for the 
physical masses of the corresponding mesons involved in the process.}. 

\subsection{Invariant amplitude}
We shall consider the generic $K_{\ell 3}$  process
\beq
K (p_K) \to \pi (p_\pi) \, \ell^+ (p_\ell) \, \nu_\ell (p_\nu) 
 .
\eeq
The invariant amplitude reads
\beqa
\lefteqn{{\cal M} = 
\frac{G_{\rm F}}{\sqrt{2}} V_{us}^{*} \, l^{\mu} \, C} \nonumber \\* 
&\times& \bigg[ f_{+}^{K \pi} (t) \, (p_K + p_\pi)_{\mu} 
+  f_{-}^{K \pi} (t) \, (p_K - p_\pi)_{\mu} \bigg]  ,
\label{basic1}
\eeqa
where 
$$ l^\mu = \bar{u} (p_\nu) \, \gamma^\mu \, (1 - \gamma_5) \, v (p_\ell) 
\, ,
\quad C = \left\{ \begin{array}{ll} 1 & \mbox{for} \ K^{0}_{\ell 3} \\
\frac{1}{\sqrt{2}} & \mbox{for} \ K^{+}_{\ell 3} \end{array} \right.  .  
$$
The expression in parentheses corresponds to the matrix element $
\langle \pi (p_\pi) | V_{\mu}^{4-i5} | K (p_K) \rangle $, expressed in
terms of the form factors $f_{\pm}^{K \pi} (t)$. The hadronic form 
factors depend on the single variable $t = (p_K - p_\pi)^2$. 

\subsection{Dalitz Plot density}

It is customary to analyze the spin-averaged decay distribution $\rho
(y,z)$ for $K_{\ell 3}$.  It depends on two variables, for which we 
choose:
\beq z = \frac{ 2 p_\pi \cdot p_K }{M_K^2} = 
\frac{2 E_{\pi}}{M_K}  ,  \quad y = \frac{ 2 p_K \cdot p_\ell }{M_K^2} =
\frac{2 E_{\ell}}{M_K}  , \eeq
where $E_\pi$ ($E_{\ell}$) is the pion (charged lepton) energy in the 
kaon rest frame, and $M_K$ indicates the mass of the decaying kaon. 
Alternatively one may also use two of the Lorentz 
invariants 
\beq
t = (p_K - p_\pi)^2   , \quad u = (p_K - p_\ell)^2   , 
\quad s = (p_\pi + p_\ell)^2 
 . 
\eeq
Then the distribution (without radiative corrections) reads 
\beqa
\rho^{(0)} (y,z) &=&  {\cal N}
\Big[ A_1^{(0)}  |f_{+}^{K \pi} (t)|^2  
+ A_2^{(0)}   f_{+}^{K \pi} (t)   f_{-}^{K \pi} (t) \nn 
&& \quad {}+ A_3^{(0)}   |f_{-}^{K \pi} (t)|^2 \Big]  ,  
\label{basic2}
\eeqa
with  
\beq
{\cal N}  = C^2 \frac{G_{\rm F}^2 \, |V_{us}|^2 M_K^5}{128 \pi^3}  , 
\quad \Gamma = \int\limits_{\cal D}  dy \, dz \ \rho^{(0)} (y,z)  .  
\eeq
Moreover, defining 
\beq
r_\ell = \frac{m_\ell^2}{M_K^2}  , \qquad r_\pi = 
\frac{M_{\pi}^2}{M_K^2}  , 
\eeq
the kinematical densities are: 
\beqa
A_1^{(0)} (y,z) &=& 4 (z + y - 1) (1 - y) 
+ r_\ell (4 y + 3 z - 3) \no \\
&&{}- 4 r_\pi + r_\ell (r_\pi - r_\ell)    , \nonumber \\
A_2^{(0)} (y,z) & = & 2 r_\ell (3 - 2 y - z + r_\ell - r_\pi)  , \nn
A_3^{(0)} (y,z) & = &  r_\ell ( 1 + r_\pi - z - r_\ell)   . 
\eeqa
In the analysis of $K_{e3}$ decays, the terms proportional to
$A_{2,3}^{(0)}$ can be neglected, being proportional to $r_e \simeq
10^{-6}$. 
The physical domain ${\cal D}$ is defined by 
\beqa \label{domain1}
2 \sqrt{r_\ell} & \leq y \leq & 1 + r_\ell - r_\pi  , \nonumber \\ 
a (y) - b (y) & \leq z \leq & a (y) + b (y)  ,  
\eeqa
where  
\beqa \label{domain2}
a (y) & = & \frac{(2 - y) \, (1 + r_\ell + r_\pi -y)}{2 ( 1 + r_\ell - y)}
 , \nonumber \\ 
b (y) & = & \frac{\sqrt{y^2 - 4 r_\ell} \, 
(1 + r_\ell - r_\pi -y)}{2 ( 1 + r_\ell - y)} ,   
\eeqa
or, equivalently,
\beqa \label{domain3}
2 \sqrt{r_\pi} & \leq z \leq & 1 + r_\pi - r_\ell  , \nonumber \\ 
c (z) - d (z) & \leq y \leq & c (z) + d (z)  ,  
\eeqa
where  
\beqa \label{domain4}
c (z) & = & \frac{(2 - z) \, (1 + r_\pi + r_\ell -z)}{2 ( 1 + r_\pi - z)}
 , \nonumber \\ 
d (z) & = & \frac{\sqrt{z^2 - 4 r_\pi} \, 
(1 + r_\pi - r_\ell -z)}{2 ( 1 + r_\pi - z)}  . 
\eeqa

\section{Structure of radiative corrections to $\mbox {\boldmath $K_{\ell 
3}$}$ decays}
\label{sec: structure} 
\renewcommand{\theequation}{\arabic{section}.\arabic{equation}}
\setcounter{equation}{0}

In decays involving charged particles, the observable quantity always
involves an inclusive sum over the {\em parent} mode and final states
with additional photons.  Therefore, when considering radiative
corrections one must include the effect of real photon radiation as
well as virtual electromagnetic corrections. Moreover, from a
theoretical point of view, only such an inclusive sum is free of
infrared (IR) divergences order by order in $\alpha$.  In the present
section we give an overview on the analysis of $K_{\ell 3}$ decays with
inclusion of radiative corrections. We show how electromagnetic 
corrections can be accounted for by using generalized form factors 
and kinematical densities, whose specific form we describe in detail 
in the following sections. 

The virtual corrections will in general produce a shift in the
invariant amplitude ${\cal M} \to {\cal M} +
\Delta {\cal M}$.  On the other hand, the radiation of 
real photons (governed by an amplitude ${\cal M}^{\gamma}$) will 
produce a shift to the decay distribution and rate.  

The virtual electromagnetic corrections do not alter the structure of
the invariant amplitude (\ref{basic1}) in terms of the form factors
$f_\pm (t)$, but change the form factors themselves. We denote by
$F_\pm (t,v)$ the full form factors including virtual electromagnetic
corrections. Note that such objects now depend also on a second
kinematical variable (this is because $F_\pm$ cannot be interpreted
anymore as matrix elements of a quark current between hadronic
states). The variable $v$ is taken as $u = (p_K - p_\ell)^2$ for
$K_{\ell 3}^+$ and $s = (p_\pi + p_\ell)^2$ for $K_{\ell 3}^0$.
We observe here that $F_\pm$ contain infrared singularities,
due to low-momentum virtual photons: we regularize them by introducing
a small photon mass $M_\gamma$.  Summarizing, the effect of virtual
corrections is to change the invariant amplitude (\ref{basic1}) as
follows:
\beq 
{\cal M} \, \big[f_{+}^{(0)} (t),f_{-}^{(0)} (t)\big] 
\to {\cal M} \, \big[ F_+ (t,v),F_- (t,v) \big]  . 
\eeq
Moreover, it is convenient to factor out of $F_\pm$ the long distance
component $\Gamma_c (v,m_\ell^2,M^2;M_\gamma)$ of loop amplitudes, 
which generates infrared and Coulomb singularities, as follows:
\beq 
F_\pm (t,v) = \left[ 1 + \frac{\alpha}{4 \pi} 
\Gamma_c (v,m_\ell^2,M^2;M_\gamma) \right] \ f_\pm (t,v)   . 
\label{factor1}
\eeq 
$\Gamma_c$ depends on the mass $m_\ell$ of the charged lepton, the
mass $M$ of the charged meson, the variable $v$ and has a logarithmic
dependence with respect to the infrared regulator $M_\gamma$.  The
definition of $\Gamma_c$ is not unique (due to possible shifts in the
infrared finite parts), and we give our deifinition in 
(\ref{Gammac}).  Once one specifies the form of $\Gamma_c$, 
(\ref{factor1}) serves also as a definition of the
structure-dependent effective form factors $f_\pm (t,v)$.  In the next
section we shall present the calculation of $f_\pm$ at next to leading
order in CHPT.

Let us now comment on the radiative amplitude ${\cal M}^\gamma$, 
and how it affects the decay distribution.   
In the intermediate stages of this  work we shall use Low's theorem 
\cite{low} to obtain the leading components of ${\cal M}^\gamma$ at 
small photon momentum. This is consistent with a full calculation 
at order $e^2 p^2$ in CHPT\footnote{At this order 
CHPT reproduces the results of Low's theorem, with $f_{+} (t) = 1 $ and 
$f_{-} (t) = 0$.}. However, it allows us to be more general in the 
derivation of the radiatively-corrected decay distribution, by including 
at intermediate steps some higher order terms in the chiral counting. 
We restore the correct chiral counting by inserting in the final expression 
for the decay rate the form factors as calculated at $\cO(p^4,e^2 p^2)$ 
in CHPT. Let us now sketch the analysis. 
The radiative amplitude can be expanded in powers of the photon energy 
$E_\gamma$, 
\beq 
{\cal M}^\gamma = {\cal M}_{(-1)}^{\gamma}
+ {\cal M}_{(0)}^{\gamma} + \cdots  , 
\label{low1}
\eeq 
where 
\beq {\cal
M}_{(n)}^{\gamma} \sim E_\gamma^n   .  
\eeq 
Gauge invariance relates ${\cal M}_{(-1)}^{\gamma}$ 
and ${\cal M}_{(0)}^{\gamma}$ to the 
non-radiative amplitude ${\cal M}$, and thus to the full form factors 
$F_{\pm}(t,v)$. 
Upon taking the square modulus and summing over spins, the radiative
amplitude generates a correction $\rho_\gamma (y,z)$ to the Dalitz
plot density of (\ref{basic2}). The observable distribution is 
now the sum 
\beq 
\rho (y,z) =  \rho^{(0)} (y,z) \, + \rho_\gamma  (y,z)   .  
\eeq
Both terms on the right hand side of this equation depend on the full
form factors $F_\pm$ and contain infrared divergences (from virtual or real
soft photons). Upon combining them, the observable density can be
written in terms of new kinematical densities $A_i$ and the effective 
form factors $f_\pm (t,v)$ defined in (\ref{factor1}), 
\beqa \label{fulldensity}
\lefteqn{\rho (y,z) = {\cal N} \, \, S_{\rm EW}(M_\rho,M_Z)}  \\*  
&& \times \Big[A_1  |f_{+} (t,v)|^2   
+ A_2  f_{+} (t,v)   f_{-} (t,v)  
+ A_3  |f_{-} (t,v)|^2 \Big] , \nonumber
\eeqa
where we have pulled out the short-distance enhancement factor 
$S_{\rm EW}$ discussed in Sect. \ref{sec: SM}.
To first order in $\alpha$, the kinematical densities are given by
\beq \label{deltas}
A_i (y,z) = A_i^{(0)} (y,z) \,  \left[ 1 + \Delta^{\rm IR} (y,z) \right] 
\,  + 
 \, \Delta_{i}^{\rm IB} (y,z)   . 
\eeq
The function $\Delta^{\rm IR} (y,z)$ arises by combining the contributions 
from $|{\cal
M}_{(-1)}^{\gamma}|^2$ and $\Gamma_{c} (v,m_\ell^2,M^2; M_\gamma)$. 
Although the individual contributions contain infrared divergences, the 
sum is  finite.  
The factors $\Delta^{\rm IB}_i (y,z)$ originate from averaging the 
remaining terms of $|{\cal M}^{\gamma}|^2$ [see (\ref{low1})]  
and are IR finite. Note that both $\Delta^{\rm IR} (y,z)$ and 
$\Delta^{\rm IB}_i (y,z)$ are sensitive to the treatment 
of the real photon emission. 
Details on these corrections are given in Sect.~\ref{sec: corr}. \\
Let us finally note that, in principle, the radiative amplitude
generates new terms in the density, proportional to derivatives of
form factors.  These terms would only arise at order $e^2 p^4$ and
higher in CHPT, and therefore we have suppressed them in
(\ref{fulldensity}).

Summarizing, the resulting Dalitz density is formally identical to the 
unperturbed one, but now one has
\beqa
f_\pm^{(0)} (t) &  \to & f_\pm (t,v)  , \nonumber \\ 
A_i^{(0)} (y,z) &  \to & A_i (y,z)  . 
\eeqa 
We want to stress here that although the structure dependent 
electromagnetic effects in $f_\pm (t,v)$ 
are due to the interplay between QCD dynamics and QED, 
the corrections of order $\alpha$ included in the densities $A_i (y,z)$ 
are universal in the sense that they are fixed by gauge invariance and 
kinematics. However they do depend on the choice of the experimental cuts 
to the photon spectrum and they exhibit a certain ambiguity due to 
possible different definitions of the function 
$\Gamma_c(v,m^2_\ell,M^2;M_\gamma)$ introduced in (\ref{factor1}) 
(we adopt the form given in (\ref{Gammac}) below). 
We suggest that these modified densities be used in the 
data analysis. This does not introduce any model dependence, and takes 
care of very long distance electromagnetic corrections. 
Such an experimental analysis of the Dalitz plot density 
could then provide valuable information on $f_\pm(t,v)$, to be confronted 
with theoretical calculations of pure QCD and QCD$+$QED effects.

\section{$\mbox{\boldmath $K_{\ell 3}$}$ amplitudes at order 
$\mbox{\boldmath $p^4$}$ and $\mbox{\boldmath $e^2 p^2$}$ 
in CHPT} 
\label{sec: CHPT}
\renewcommand{\theequation}{\arabic{section}.\arabic{equation}}
\setcounter{equation}{0}

In this section we analyze the form factors relevant for $K_{\ell 3}$
decays in the framework of CHPT, including electromagnetic corrections at 
order $e^2 p^2$.  As for the non-electromagnetic part, we shall give
the results at order $p^4$ in standard CHPT, including the effect of
{\em strong} isospin breaking ($m_u \neq m_d$). In principle, the order
$p^6$ corrections can be easily included in our formalism, in order to
make a more accurate phenomenological analysis.

In Sect. \ref{sec: structure} we have introduced the effective form
factors $f_{\pm} (t,v)$, from which the long distance virtual photon
effects have been removed [see (\ref{factor1})].
We also stressed that in order to give a meaningful definition 
of such objects one has to specify the function 
$\Gamma_c(v,m_\ell^2,M^2;M_\gamma)$, arising in perturbation 
theory from photonic loop contributions to the form factor $F_+ (t,v)$:
\beqa \label{Gammac}
\lefteqn{\Gamma_c (v,m_\ell^2,M^2;M_\gamma)  =} \nonumber \\*  
&& 2 M^2 Y \, {\cal C} (v,m_\ell^2,M^2)  
+ 2  \log \frac{M m_\ell}{M_\gamma^2} 
\bigg(1 + \frac{X Y 
\log X}{\sqrt{R} 
(1 - X^2)}  \bigg)  . \nn
&&
\eeqa
In Appendix \ref{appB} we define the auxiliary variables $X$, $R$,
$Y$, as well as the function ${\cal C} (v,m_\ell^2,M^2)$. 
$\Gamma_c$ encodes the leading Coulomb interaction between the charged
particles involved in the decay ($K^+$ and $\ell^+$ for $K_{\ell
3}^+$, $\pi^-$ and $\ell^+$ for $K_{\ell 3}^0$).  It is IR divergent
and singular when the two charged particles are relatively at rest. In
the case of $K^+$ decays, this singularity is outside the physical
region, while it occurs on its boundary for the $K^0$ decay, implying
a larger overall effect.  It is convenient to write the form factors
in a form suitable for a resummation of (potentially large) IR and
Coulomb effects.  At the same time we want this representation to
reproduce the correct CHPT result at order $e^2 p^2$. We achieve this
by writing the full form factors as follows, in terms of the effective
form factors:
\beqa
F_{\pm}(t,v) &=& \Big[1\,+\,\frac{\alpha}{4\pi}\,
\Gamma_c(v,m_{\ell}^2,M^2 ; M_\gamma)\Big]\,f_{\pm}(t,v) \ , \nn
f_{\pm}(t,v) &=& \wt{f}_{\pm} (t) \,+\, f_{\pm}^{\mbox{\small{EM-loc}}}\,
+\, f_{\pm}^{\mbox{\small{EM-loop}}}(v)
\, .
\label{eff}
\eeqa
Let us now comment on each term appearing in (\ref{eff}):
\begin{itemize}
\item $\wt{f}_\pm (t)$ represent the pure QCD contributions to the 
form factors (in principle at any order in the chiral expansion), plus 
the electromagnetic contributions up to order $e^2 p^2$ 
generated by the non-derivative Lagrangian defined in (\ref{Leff})
(proportional to $Z$). Diagrammatically they arise from purely mesonic 
graphs. In the definition of $\wt{f}_{+}^{K^+ \pi^0} (t)$, 
we have included also the electromagnetic counterterms relevant 
to $\pi^0$--$\eta$ mixing [see (\ref{eps4S}) and (\ref{eps4EM}) below].  
\item $f_{\pm}^{\mbox{\small{EM-loc}}}$  represent the local effects 
of virtual photon exchange. 
\item $f_{\pm}^{\mbox{\small{EM-loop}}}(v)$ represent the non-local 
photonic loop contribution (once the Coulomb term has been removed). 
We have checked that in all relevant cases
$f_{\pm}^{\mbox{\small{EM-loop}}}(v)$ are smooth functions of the
variable $v$ in the physical region, allowing one to perform a linear
expansion in the Dalitz variables $y,z$.
\end{itemize}
Each term in the above decomposition is scale-independent.
We now give the full expressions for the form factor compponents, 
in terms of loop functions defined in Appendix~\ref{appA}. 

\subsection{The form factors $\mbox{\boldmath $f_{\pm}^{K^+ \pi^0} 
(t,u)$}$}

For $f_{\pm}^{K^+\pi^0}$ the mesonic contribution is given by: 
\begin{eqnarray}
 \wt{f}_{+} (t) & = & 1 + \sqrt{3} \,  \Big( \ve^{(2)} + 
\ve^{(4)}_{\rm S} + \ve^{(4)}_{\rm EM} \Big) \nonumber \\*
&& {} +  \frac{1}{2} H_{K^+ \pi^0} (t) +  
 \frac{3}{2} H_{K^+ \eta} (t) +  H_{K^0 \pi^-} (t)    \nonumber \\*
&& {} +  \sqrt{3} \, \ve^{(2)} \Bigg[
 \frac{5}{2} H_{K \pi} (t) + \frac{1}{2} H_{K \eta} (t) \Bigg]  . 
\label{ff1}
\end{eqnarray}
This expression is essentially the pure QCD form factor at $\cO(p^4)$ 
\cite{gl852}, except for the inclusion of electromagnetic contributions 
to the meson masses and for the additional contribution  
$\ve^{(4)}_{\rm EM}$, due to $\pi^0$--$\eta$ mixing at $\cO(e^2 p^2)$ 
\cite{nr95}. 
The function $H_{PQ}(t)$ \cite{gl852,GL85} is reported in 
Appendix \ref{appA}, and the leading order $\pi^0$--$\eta$ mixing 
angle $\ve^{(2)}$ has already been given in (\ref{epsilon}). 
The remaining quantities in (\ref{ff1}) are given by
\beqa \label{eps4S}
\lefteqn{\ve^{(4)}_{\rm S} = 
- \frac{2 \, \ve^{(2)}}{3 (4 \pi F_0)^2 (M_{\eta}^2 - M_{\pi}^2)}} \nonumber 
\\* 
& \times &  \bigg\{ (4 \pi)^2 \, 64 \left[3 L_7 + 
L_8^r (\mu) \right] 
(M_K^2 - M_\pi^2)^2 
\nn
&& {} -  M_\eta^2 (M_K^2 - M_\pi^2) \log \frac{M_\eta^2}{\mu^2}
 +  M_\pi^2 (M_K^2 - 3 M_\pi^2) \log \frac{M_\pi^2}{\mu^2}  \nonumber \\
&& {} - 2 M_K^2 (M_K^2 - 2 M_\pi^2) \log \frac{M_K^2}{\mu^2} 
- 2 M_K^2 (M_K^2 - M_\pi^2) \bigg\}  , \nn
&&   
\eeqa
and \footnote{ The last $-1$ in Eq. (5.21) of \cite{nr95} should be
replaced by $+ 1$; the preceding formula (5.20) is correct.}
\beqa \label{eps4EM}
\lefteqn{\ve^{(4)}_{\rm EM} =   
\frac{2 \, \sqrt{3} \, \alpha \, M_K^2}{108 \, \pi \, (M_\eta^2 
-M_\pi^2)} } \no \\*
& \times & \bigg\{ 2 (4 \pi)^2 \Big[
-6  K_3^r (\mu) + 3 K_4^r (\mu)  
+ 2 K_5^r (\mu) + 2 K_6^r (\mu) \Big]  \nonumber \\
&& {} -  9 Z \left(\log \frac{M_K^2}{\mu^2} + 1 \right) \bigg\}  . 
\eeqa  
They are related to the one-loop off-diagonal element of the squared mass 
matrix in the $\pi^0$--$\eta$ sector \cite{nr95} by
\beq
\ve^{(4)}_{\rm S} + \ve^{(4)}_{\rm EM} = - \frac{M^2_{\hat{\pi}^0 
\hat{\eta}}}{(M^2_\eta - M^2_\pi)} .
\eeq
The analogous contribution for the $f_{-}$ form factor is given by: 
\beqa
\wt{f}_{-}(t) &=&
\frac{4}{F_0^2}(1\,+\,\frac{\varepsilon^{(2)}}{\sqrt{3}})(M_K^2-M_\pi^2) 
\nonumber\\
& \times & 
\Big[ L_5^r(\mu)\,-\,\frac{3}{256\pi^2}\,\ln\frac{M_{K^\pm}^2}{\mu^2} \Big] 
\nonumber\\
&-&\,\frac{1}{128\pi^2F_0^2}\,\Big[
(3+\sqrt{3}\,\varepsilon^{(2)})M_{\eta}^2\ln\frac{M_{\eta}^2}{M_{K^\pm}^2}\,
\nonumber\\
&+&\,
2(3-\sqrt{3}\,\varepsilon^{(2)})M_{K^0}^2\ln\frac{M_{K^0}^2}{M_{K^\pm}^2}
\nonumber\\
&-&\,
2(3-\sqrt{3}\,\varepsilon^{(2)})M_{\pi^\pm}^2\ln
\frac{M_{\pi^\pm}^2}{M_{K^\pm}^2}\,
\nonumber\\
&+& (1+3\sqrt{3}\,\varepsilon^{(2)})M_{\pi^0}^2\ln
\frac{M_{\pi^0}^2}{M_{K^\pm}^2} \Big]
\nonumber\\
&+&
\,\sum_{PQ}
\Big\{ 
\Big[ a_{PQ}(t) + \frac{\Delta_{PQ}}{2t} b_{PQ}(t) \Big] K_{PQ}(t) 
\nonumber\\
&+& b_{PQ}(t)\,\frac{F_0^2}{t} H_{PQ} \Big\}  \, . 
\label{fminustildep}
\eeqa
The sum in the last line runs over meson pairs occurring in loop
diagrams.  The functions $K_{PQ} (t)$ and $H_{PQ} (t)$ \cite{GL85}
are reported in Appendix \ref{appA}, while the relevant coefficients
$a_{PQ} (t)$ and $b_{PQ} (t)$ are listed in Appendix \ref{appC}.

The local electromagnetic contributions are given by: 
\beqa
f_{+}^{\mbox{\small{EM-loc}}} &=& 4\pi\alpha \Big[
2K_{12}^r(\mu) - \frac{8}{3} X_1 - \frac{1}{2} \wt{X}_6^r(\mu) 
\nonumber\\
&-& \frac{1}{32\pi^2} \Big(
3 + \ln\frac{m_{\ell}^2}{M_{K^\pm}^2} + 3\ln\frac{M_{K^\pm}^2}{\mu^2}
\Big)\Big] 
\,, \label{factor2}
\eeqa
\beqa
f_{-}^{\mbox{\small{EM-loc}}} &=& 8\pi\alpha \Big[
2K_3^r(\mu) - K_4^r(\mu) - \frac{1}{3}\big(K_5^r(\mu)+K_6^r(\mu)\big) 
\nonumber\\
&+& X_1 - X_2^r(\mu) + X_3^r(\mu)
\nonumber\\
&-&\frac{1}{32\pi^2}\,\Big(
1 - 2\ln\frac{m_{\ell}^2}{M_{K^\pm}^2} - (3 + 
\frac{5Z}{2})\ln\frac{M_{K^\pm}^2}{\mu^2} \Big)\Big] \, .
\nonumber\\
&&
\eeqa

Finally, the loop contributions are given by 
\beq
f_{\pm}^{\mbox{\small{EM-loop}}}(u) \,=\,\frac{\alpha}{4\pi}\,\Big[
\Gamma_1(u,m_{\ell}^2,M_K^2) \,
\pm\, \Gamma_2(u,m_{\ell}^2,M_K^2)
\Big]\,,
\label{fplusEM-LOOP}
\eeq
with $\Gamma_{1,2} (v,m_{\ell}^2,M^2)$ given in Appendix \ref{appB}.

\subsection{The form factors $\mbox{\boldmath $f_{\pm}^{K^0 \pi^-} 
(t,s)$}$}

In the case of $f_{\pm}^{K^0\pi^+}$,
the mesonic loop contribution is given by 
\begin{eqnarray}
 \wt{f}_{+} (t) & = & 1 + \frac{1}{2} H_{K^+ \pi^0} (t) +  
 \frac{3}{2} H_{K^+ \eta} (t) +  H_{K^0 \pi^-} (t)    \nonumber \\*
&&{} + \sqrt{3} \, \ve^{(2)} \left[ H_{K \pi} (t) -  
H_{K \eta} (t) \right]  \ ,  
\end{eqnarray}
and 
\beqa
\wt{f}_-(t) &=&
\frac{4}{F_0^2}(1\,+\,\frac{2 \, \varepsilon^{(2)}}{\sqrt{3}})(M_K^2-M_\pi^2)
\nn
&\times & \Big[ 
L_5^r(\mu)\,-\,\frac{3}{256\pi^2}\,\ln\frac{M_{\pi^\pm}^2}{\mu^2} \Big]
\nonumber\\
&-&
\frac{1}{128\pi^2F_0^2}\,\Big[
(3+2\sqrt{3}\,\varepsilon^{(2)})M_{\eta}^2\ln\frac{M_{\eta}^2}{M_{\pi^\pm}^2}
\nn
&+& 
2M_{K^0}^2\ln\frac{M_{K^0}^2}{M_{\pi^\pm}^2} \,-\,
(3+2\sqrt{3}\,\varepsilon^{(2)})M_{\pi^0}^2\ln\frac{M_{\pi^0}^2}{M_{\pi^\pm}^2}
\Big]
\nonumber\\
&+& 
\sum_{PQ}
\Big\{ 
\Big[ c_{PQ}(t) + \frac{\Delta_{PQ}}{2t} d_{PQ}(t) \Big] K_{PQ}(t) 
\nonumber\\
&+& d_{PQ}(t)\,\frac{F_0^2}{t} H_{PQ} \Big\}  \, .
\label{fminustildez}
\eeqa
As previously, the sum runs over meson pairs occurring in loop
diagrams. The coefficients $c_{PQ} (t)$ and $d_{PQ} (t)$, for the
relevant meson pairs, are reported in Appendix \ref{appC}.

The local electromagnetic terms are given by:
\beqa
f_{+}^{\mbox{\small{EM-loc}}} &=& 4\pi\alpha \Big[
2K_{12}^r(\mu) \, + \, \frac{4}{3} X_1 - \frac{1}{2} \wt{X}_6^r(\mu) 
\nn
&-& 
\frac{1}{32\pi^2} \Big( 3 + \ln\frac{m_{\ell}^2}{M_{\pi^\pm}^2} + 
3\ln\frac{M_{\pi^\pm}^2}{\mu^2} \Big)\Big]
\, ,
\eeqa
\beqa
f_{-}^{\mbox{\small{EM-loc}}} &=& 8\pi\alpha \Big[
 - \frac{1}{3}\big(K_5^r(\mu)+K_6^r(\mu)\big) 
\nn 
&+&
X_1 + X_2^r(\mu) - X_3^r(\mu)
\nonumber\\
&+&\,\frac{1}{32\pi^2}\,\Big(
1 - 2\ln\frac{m_{\ell}^2}{M_{\pi^\pm}^2} - 
(3 - \frac{Z}{2})\ln\frac{M_{\pi^\pm}^2}{\mu^2} \Big)\Big]\, . \nn
&& 
\eeqa

Finally, the loop contributions are: 
\beq
f_{\pm}^{\mbox{\small{EM-loop}}}(s) \,=\,\frac{\alpha}{4\pi}\,\Big[
\Gamma_2(s,m_{\ell}^2,M_{\pi}^2) \,\pm\, \Gamma_1(s,m_{\ell}^2,M_{\pi}^2)
\Big]
\,.
\eeq

\subsection{The effective form factors for $K_{e 3}$ decays}
In the analysis of $K_{e3}$ decays (where only $F_+$ is observable), 
in order to avoid unnecessary complications we can write  
\beqa
\lefteqn{F_+ (t,v)  = }  \nonumber \\*
&& f_+(t)
\left\{ 1 + \frac{\alpha}{4 \pi} \Gamma_c (v,m_\ell^2,M^2;M_\gamma) + 
f_{+}^{\mbox{\small{EM-loop}}}(v) \right\} ,
\nonumber \\*
&&
\eeqa
with
\beq 
f_{+} (t) = \wt{f}_+ (t) +  f_{+}^{\mbox{\small{EM-loc}}} \ . 
\eeq
Exploiting the ambiguity in the definition of the function $\Gamma_c$,
here we choose to shift a smooth function of $v$ from the form factor
to $\Gamma_c$ itself. The advantage is that the effective form factor
becomes a function of the single variable $t$.  This trick is not
possible in the treatment of $K_{\mu 3}$ decays, as the loop
contributions $f_{+}(v)$ and $f_{-}(v)$ are different.

\section{Corrections to kinematical densities}
\label{sec: corr}
\renewcommand{\theequation}{\arabic{section}.\arabic{equation}}
\setcounter{equation}{0}

As an illustration of the general considerations of Sect. 
\ref{sec: structure}, 
we present here in detail a possible treatment of the contribution of the 
real photon emission process
\beq
K^+(p_K) \to \pi^0 (p_\pi) \,  \ell^+ (p_\ell) \, 
\nu_\ell (p_\nu) \, \gamma (p_\gamma)  .
\eeq
Following the procedure proposed by Ginsberg \cite{gin67}, we define the 
kinematical variable
\beq \label{x}
x = (p_\nu + p_\gamma)^2 = (p_K - p_\pi - p_\ell)^2  . 
\eeq
A possible choice of the purely radiative part of the decay distribution 
(corresponding to a specific analysis of the experimental data) is to 
accept all pion and charged lepton energies
within the whole $K_{e3}$ Dalitz plot $\D$ and all kinematically 
allowed values of the Lorentz invariant $x$ defined in (\ref{x}). 
(The variable $x$ determines the angle between the pion and lepton 
momentum for given energies $E_\pi$, $E_\ell$.)
This translates into: 
\beqa \label{rhogamma}
\lefteqn{\rho_\gamma (y,z) =  \frac{M_K}{2^{12}\pi^5} 
\int\limits_{M^2_\gamma}^{x_{\rm max}} 
dx} \nonumber \\*
& \times & \! \frac{1}{2 \pi} \! \int \! \frac{d^3 p_\nu}{p_\nu^0} 
\frac{d^3 p_\gamma}{p_\gamma^0} 
\delta^{(4)}(p_K-p_\pi-p_\ell-p_\nu-p_\gamma) \! 
\sum_{\rm pol}  |{\cal M}^{\gamma}|^2   , \nonumber \\*
&&
\eeqa 
with
\beqa \label{xmax} 
x_{\rm max} &=& M_K^2 
\bigg\{1 + r_\pi + r_\ell - y - z  \nn
&&{}+ \frac{1}{2} \Big[y z  +  \sqrt{(y^2 - 4 r_{\ell})(z^2 - 4 r_\pi)} 
\Big] \bigg\} . 
\eeqa
In (\ref{rhogamma}) we have extended the integration over the whole range 
of the invariant mass of the unobserved $\nu_\ell \, \gamma$ system.
The integrals occurring in (\ref{rhogamma}) have the general form 
\cite{gin67}
\beqa \label{Imn}
\lefteqn{I_{m,n}(p_1,p_2;P,M_{\gamma}) :=} \nonumber \\* 
&& \frac{1}{2 \pi} \int 
\frac{d^3 q}{q^0} \frac{d^3 k}{k^0} 
\frac{\delta^{(4)}(P-q-k)} 
{(p_1 \cdot k 
+M_{\gamma}^2/2)^m (p_2 \cdot k + M_{\gamma}^2/2)^n}  . \nonumber \\*
&&
\eeqa
The results for these integrals in the limit $M_\gamma = 0$ can be found 
in the Appendix of \cite{gin67}. (We have checked these expressions.)
Using the definition (\ref{Imn}), the  radiative decay 
distribution (\ref{rhogamma}) can be written as 
\cite{gin67}
\beqa \label{decomp}
\lefteqn{\rho_\gamma (y,z) =} \nonumber \\* 
&& \frac{\alpha}{\pi} \Bigg[ \rho^{(0)} (y,z) I_0(y,z;M_\gamma) 
+\frac{G_{\rm F}^2 |V_{us}|^2 |f_{+}|^2 M_K}{128 \pi^3} \nn
&& \times \! \! \int\limits_0^{x_{\rm max}} \! \! dx 
\sum_{m,n} c_{m,n} I_{m,n}(p_\ell,p_K;p_K-p_\pi-p_\ell,0) \Bigg]  ,
\nonumber \\*
&&
\eeqa
where the infrared divergences are now confined to
\beqa \label{I0}
\lefteqn{I_0(y,z;M_\gamma) =}  \nn
&& \int\limits_{M^2_{\gamma}}^{x_{\rm max}} dx 
\Big[- 2 p_K \cdot p_{\ell} 
I_{1,1}(p_\ell,-p_K;p_K-p_\pi-p_\ell;M_\gamma) 
\nn[-14pt]
&& \qquad \quad {} - M_K^2 
I_{0,2}(p_\ell,-p_K;p_K-p_\pi-p_\ell;M_\gamma)
\nn
&& \qquad \quad {}- m_{\ell}^2 
I_{2,0}(p_\ell,-p_K;p_K-p_\pi-p_\ell;M_\gamma) \Big] . 
\eeqa
The explicit form of the function $I_0$ can be found in 
Eq. (27) of \cite{gin67}.
The coefficients $c_{m,n}$ were given in Eq. (19) of \cite{gin67}. Note 
however the misprint for the values of $c_{-1,0}$ and $c_{1,-2}$ (see 
Erratum of \cite{gin67}). (We have also checked this list of 
coefficients.) 

The function $\Delta^{\rm IR}$ introduced in (\ref{deltas}) can now be 
related to $I_0$ by
\beq
\Delta^{\rm IR}(y,z) = \frac{\alpha}{\pi} \left[ I_0(y,z;M_\gamma)
+ \frac{1}{2} \Gamma_{\ell}(u,m^2_\ell,M_K^2;M_\gamma) \right] \ , 
\label{deltaIR}
\eeq
where
\beq
\Gamma_\ell = \left\{ \begin{array}{ll} 
\Gamma_c  + \frac{4 \pi}{\alpha} f_{+}^{\mbox{\small{EM-loop}}} 
& \quad \mbox{for} \ \ell = e \\
\Gamma_c   
& \quad \mbox{for} \ \ell = \mu \end{array} \right.  .  
\eeq

An analytic expression of the integral occurring in (\ref{decomp}) 
 was given in 
Appendix B of \cite{gin70} in terms of the quantities $U_i$:
\beq
\int\limits_0^{x_{\rm max}} dx \: \sum_{m,n} c_{m,n} I_{m,n} = 
\sum_{i=0}^7 U_i
 .
\eeq
Note that the quantity $J_9(i)$ given in Eq. (A9) of \cite{gin70} (which 
is needed for the evaluation of $U_7$) contains two crucial mistakes:
the plus-sign in the last line of (A9) should be replaced by a minus-sign, 
and $|\beta_i^{\rm max}|$ at the end of the first line of (A9) should 
simply read $\beta_i^{\rm max}$. The second error is irrelevant for $\ell 
= e$ 
but has disastrous consequences in a certain part of the Dalitz plot for 
$\ell = \mu$. To the best of our knowledge none of these errors has been 
reported in an Erratum of \cite{gin70}. 

The functions $\Delta_i^{\rm IB}$ introduced in (\ref{deltas}) can now be 
written as
\beqa \label{DeltaIB}
\Delta_1^{\rm IB} &=& \frac{2 \alpha}{\pi M_K^4} \sum_{i=0}^7 U_i 
\Big|_{\xi=0}  , \no \\
\Delta_2^{\rm IB} &=& \frac{2 \alpha}{\pi M_K^4} \sum_{i=0}^7 U_i 
\Big|_{\xi}  , \no \\
\Delta_3^{\rm IB} &=& \frac{2 \alpha}{\pi M_K^4} \sum_{i=0}^7 U_i 
\Big|_{\xi^2}  ,
\eeqa
where the symbol $\xi$ used in \cite{gin70} stands for the ratio $f_- / 
f_+$.
 
\section{Application to  $\mbox{\boldmath $K^{+}_{e3}$}$ decay}
\label{sec: applic}
\renewcommand{\theequation}{\arabic{section}.\arabic{equation}}
\setcounter{equation}{0}

In this section we provide an illustrative analysis of the
$K^{+}_{e3}$ decay with inclusion of isospin-breaking and radiative
corrections.  The aim here is to illustrate how one might proceed in
order to extract $\vert V_{us} \vert$ rather than to give final
numbers, which would anyhow still depend on the size of the hitherto
unknown two-loop strong interaction contributions.  Once they become
known, these corrections could then easily be incorporated into the
analysis presented here.

As usual we provide a linear expansion for the effective form factor 
$f_{+}^{K^+ \pi^0} (t)$, which reads:
\beq \label{ex1}
f_{+}^{K^+ \pi^0} (t) = f_{+}^{K^+ \pi^0} (0) \ \bigg[ 1 +  
 \frac{t}{M_{\pi^{\pm}}^2} \wt{\lambda}_+ \bigg]  , 
\eeq
with
\beq
f_{+}^{K^+ \pi^0} (0) = 
 \wt{f}_{+} (0) +  f_{+}^{\mbox{\small{EM-loc}}}  \ , 
\eeq
and
\beq
\frac{\wt{\lambda}_+}{M_{\pi^{\pm}}^2} = 
\frac{d \wt{f}_{+} (t)}{d t} \bigg|_{t=0}  . 
\eeq
Using the linear expansion for the form factor 
$f^{K^+ \pi^0} (t)$, the decay rate
\beqa 
\lefteqn{\Gamma(K^+ \to \pi^0 e^+ \nu_e (\gamma)) :=} \nonumber \\*
&& \Gamma(K^+ \to \pi^0 e^+ \nu_e) +
\Gamma(K^+ \to \pi^0 e^+ \nu_e \gamma) 
\eeqa
can be expressed as 
\beqa
\lefteqn{ \Gamma(K^+ \to \pi^0 e^+ \nu_e (\gamma)) =} \nonumber \\*
&&  {\cal N} \, \,  S_{\rm EW} (M_\rho,M_Z) \, 
\,  
|f_{+}^{K^+ \pi^0} (0)|^2  \, \, I(\wt{\lambda}_+)  ,
\label{ex2}
\eeqa
where 
\beqa
I (\wt{\lambda}_+) &=&  \int\limits_{\cal D}  dy \, dz \, A_1 (y,z) 
\bigg[ 1 + \frac{t}{M_{\pi^{\pm}}^2} \wt{\lambda}_+ \bigg]^2 \nonumber \\
&=& a_0 + a_1 \, \wt{\lambda}_+  + a_2 \,  \wt{\lambda}_{+}^2  . 
\label{ex3}
\eeqa
In (\ref{ex2}) we have pulled out the short distance enhancement factor 
$S_{\rm EW} (M_\rho,M_Z)$ from the low energy constant $X_6^r (\mu)$ [see 
(\ref{SEW})]. 

In order to extract $|V_{us}|$ we have to provide a theoretical
estimate of the form factor at $t=0$ and the phase-space integral.

\subsection{Numerical estimate of $f_{+}^{K^ + \pi^0} (0)$}

In our numerical analysis, we use the physical meson masses \cite{PDG}. 
We now describe the input used for the other quantities occurring 
in the form factor expansion. 

For the mixing angle defined in (\ref{epsilon}), 
we are using \cite{Leutwyler96}:
\beq \label{epsnum} 
\ve^{(2)} = (1.061 \pm 0.083) \times 10^{-2} \ . 
\eeq

The combination of coupling constants 
$3L_7 + L^r_8(M_\rho)$ entering in (\ref{eps4S})
can be determined from two observables \cite{GL85}:
the deviation from the Gell-Mann--Okubo Mass formula,
\beq
\Delta_{\rm GMO} = \frac{4M_K^2-M^2_\pi-3M^2_\eta}{M^2_\eta-M^2_\pi} ,
\eeq
which is well under control, and a quantity $\Delta_{\rm M}$ related to 
the $\cO (p^4)$ contributions for the ratio
\beq
\frac{M^2_K}{M^2_\pi} = \frac{m_s+\wh{m}}{2\wh{m}} (1+\Delta_{\rm M}) .
\eeq
Combining Eqs. (10.10) and (10.11) of \cite{GL85}, one finds
\beqa \label{COMB}
\lefteqn{3L_7 + L^r_8(M_\rho) =} \nonumber \\* 
&& F^2_0 \bigg\{ -\frac{\Delta_{\rm GMO}}{24(M^2_\eta - M^2_\pi)} 
+\frac{\mu_\eta - \mu_\pi - \Delta_{\rm M}}
{32(M^2_K - M^2_\pi)}  \nn 
&& \qquad {}+\frac{3 M^2_\eta \mu_\eta + M^2_\pi \mu_\pi - 4 M^2_K \mu_K}
{12(M^2_\eta - M^2_\pi)^2} \bigg\} ,
\eeqa 
where
\beq
\mu_P = \frac{M^2_P}{(4 \pi F_0)^2} 
\log \frac{M_P}{M_\rho} .
\eeq
Using
\beq
\Delta_{\rm M} = 0.065 \pm 0.065
\eeq
from Leutwyler's analysis \cite{Leutwyler96}, (\ref{COMB}) implies
\beq \label{COMBnum}
3L_7 + L^r_8(M_\rho) = (-0.33 \pm 0.08) \times 10^{-3} ,
\eeq
with a relatively small error. (Allowing also for higher order corrections 
one might prefer \cite{EMNP2000} a more generous bound of, say, 
$|\Delta_{\rm M}| \leq 0.2$ corresponding to
$3L_7 + L^r_8(M_\rho) = (-0.25 \pm 0.25) \times 10^{-3}$.) 

The relevant combination of electromagnetic coupling constants appearing 
in (\ref{eps4EM}) has been estimated by Bijnens and Prades \cite{bp97}. 
For our numerical analysis we add here an error of the typical size 
$1/(4 \pi)^2 \simeq 6.3 \times 10^{-3}$:
\beqa \label{bpnum}
\wh{K}^r(M_\rho) &:=& (- 6 K_3 + 3 K_4 + 2 K_5 + 2 K_6 )^r (M_\rho) \nn
&=& (5.7 \pm 6.3) \times 10^{-3} .
\eeqa
In this way we find the following mesonic contributions to the form factor: 
\beqa 
\wt{f}_{+} (0) &=& 1.0002 
\bigg( 1 + 0.0228 \cdot \frac{\ve^{(2)} - 1.06 \times 10^{-2}}
{1.06 \times 10^{-2}} \nn
&& \qquad {}+ 0.0055 \cdot 
\frac{3L_7 + L_8^r (M_\rho) + 0.33 \times 10^{-3}}{-0.33 \times 10^{-3}}
\nn
&& \qquad {}+ 0.0002 \cdot \frac{\wh{K}^r (M_\rho) - 5.7 
\times 
10^{-3}}{5.7 \times 10^{-3}} \bigg)
\nn 
&=& 1.0002 \pm 0.0018 \pm 0.0013 \pm 0.0002 \nn
&=& 1.0002 \pm 0.0022  . 
\eeqa
In the last line the three individual errors have been added in 
quadrature. The main contribution to the final error comes from the 
uncertainties in (\ref{epsnum}) and (\ref{COMBnum}). The electromagnetic 
contribution to 
$\pi^0 - \eta$ 
mixing, entering through (\ref{eps4EM}), has very little numerical 
impact on the central value and the uncertainty. 

For the coupling constant $K_{12}$ entering in the purely electromagnetic 
part (\ref{factor2}) we use a value extracted from the work of Moussallam 
\cite{moussallam}:
\beq
K^r_{12} (M_\rho) = (-4.0 \pm 0.5) \times 10^{-3} .
\eeq
For the (unknown) ``leptonic'' constants we resort to the usual 
bounds suggested by dimensional analysis:
\beq
|X_1|, \ |\wt{X}^r_6(M_\rho)| \leq 6.3 \times 10^{-3} . 
\eeq
Eventually we find: 
\beqa 
f_{+}^{\mbox{\small{EM-loc}}} &=&
 0.0032 - 0.0007 \cdot \frac{K_{12}^r (M_\rho) + 4 \times 10^{-3}}
{-4 \times 10^{-3}} 
\nonumber \\
&-&  0.0015 \cdot \frac{X_1}{6.3 \times 10^{-3}}  - 0.0003 \cdot  
\frac{\wt{X}_6^r (M_\rho)}{6.3 \times 10^{-3}} 
\nn
&=&  0.0032 \pm 0.0016  .
\eeqa
In this contribution, the sizeable relative uncertainty is almost 
exclusively due to the poor present knowledge of $X_1$. Despite this, 
in the final result for $f_{+}^{K^+ \pi^0} (0)$ this is an effect 
of $0.16 \%$. 

Combining the results given above we obtain the final value: 
\beq \label{fplusnum}
f_{+}^{K^+ \pi^0} (0) = 1.0034 \pm 0.0027  .
\eeq

\subsection{The phase space factor}

The theoretical prediction for the slope parameter is determined by the 
size of the low-energy constant $L^r_9$. With
\beq
L^r_9 (M_\rho) = (6.9 \pm 0.7) \times 10^{-3} 
\eeq
we find 
\beqa 
\wt{\lambda}_{+} & = & 0.0328  
 + 0.0321 \cdot \frac{L_9^r (M_\rho) - 6.9 \times 10^{-3}}{6.9 \times 
10^{-3}} 
\nonumber \\
&=& 0.0328 \pm 0.0033 . 
\eeqa
Here one can note the relatively large uncertainty induced by $L_9^r
(M_\rho)$. This suggests to use the measured value for the slope when 
evaluating the phase space integral. 

\begin{table}
\caption{\label{table1} Coefficients entering the phase space integral}
\vspace{0.2cm}
\begin{center}
\begin{tabular}{|l|c|c|c|}
\hline
 & $a_0$ & $a_1$ & $a_2$ \\ \hline 
$\alpha = 0$  & $0.09653  $ & $0.3337 $ & $0.4618$ \\ \hline 
$\alpha \neq 0$  & $0.09533  $ & $0.3287 $ & $0.4535$ \\ \hline 
\end{tabular}
\end{center}
\end{table}

The numerical coefficients $a_{0,1,2}$ entering in the phase space 
expression (\ref{ex3})
are reported in Table \ref{table1}. The numbers given here correspond 
to the specific prescription for the treatment of real photons described 
in the previous section: accept all pion and charged lepton energies
within the whole $K_{e3}$ Dalitz plot $\D$ and all kinematically 
allowed values of the Lorentz invariant $x$ defined in (\ref{x}). 
(The variable $x$ determines the angle between the pion and lepton 
momentum for given energies $E_\pi$, $E_\ell$.)

The inclusion of radiative corrections in this channel tends to give a
negative shift to the phase space term.  Assuming $\wt{\lambda}_+ =
0.030$, and evaluating (\ref{ex3}) with $a_{0,1,2}$ corresponding to
$\alpha \neq 0$ and $\alpha = 0$ (see Table~\ref{table1}), one can
check that radiative corrections induce in $I (\wt{\lambda}_+)$ a 
negative shift of $1.27 \%$.

\subsection{Illustrative extraction of $|V_{us}|$}

The Kobayashi--Maskawa matrix element $|V_{us}|$ can be extracted from the 
$K_{e3}$ decay parameters by
\beq \label{Vus}
|V_{us}| = \frac{16 \, \, \pi^{3/2} \, \, 
\Gamma(K^+ \to \pi^0 e^+ \nu_e (\gamma))^{1/2}}
{G_{\rm F} \, \, M_{K^{\pm}}^{5/2} \, \, S_{\rm EW}(M_\rho,M_Z)^{1/2} \, 
\, |f_+^{K^+ \pi^0}(0)| \, \, I(\wt{\lambda}_+)^{1/2}} .
\eeq
The application of this formula requires, of course, experimental numbers 
for $\Gamma(K^+ \to \pi^0 e^+ \nu_e (\gamma))$ and $\wt{\lambda}_+$
where the emission of real photons has been taken into account in accordance 
with the prescription given above. As such a consistent treatment of the 
radiative corrections cannot be guaranteed for all the experimental 
data included in the present values \cite{PDG} of the $K_{e3}$ decay 
parameters given by the Particle Data Group (PDG), the following 
analysis should only be regarded as a preliminary illustration 
of the numerics. An up-to-date determination of $|V_{us}|$ with the 
highest possible precision will then be the task of future high statistics 
$K_{e3}$ measurements, and will require the consistent inclusion of 
${\cal O} (p^6)$ terms in the form factor. 

Using \cite{PDG}
\beqa \label{Ke3par}
\Gamma(K^+ \to \pi^0 e^+ \nu_e (\gamma))
&=& (2.56 \pm 0.03) \times 10^{-15} {\rm MeV} , 
\nn
\wt{\lambda}_+ &=& 0.0276 \pm 0.0021
\eeqa
for the $K_{e3}$ parameters, $S_{\rm EW}(M_\rho, M_Z) = 1.0232$ for the 
short-distance enhancement factor \cite{MS93} (leading logarithmic and QCD 
corrections included), and the $\cO(p^4)$ prediction (\ref{fplusnum}) for 
$f_{+}^{K^+ \pi^0} (0)$, 
we find
\beqa
|V_{us}| &=& 0.2173 \pm 0.0013 \pm 0.0008 \pm 0.0006 \nn
         &=& 0.2173 \pm 0.0016 ,
\eeqa 
where the errors correspond to
\beqa
\Delta |V_{us}| &=& |V_{us}| \left( \pm \frac{1}{2} \frac{\Delta
\Gamma}{\Gamma} \pm
0.05 \cdot \frac{\Delta \wt{\lambda}_+}{\wt{\lambda}_+} \pm
\frac{\Delta f_{+}(0)}{f_{+}(0)} \right) \nn
&=& |V_{us}| \big( \pm 0.6 \% \pm 0.4 \% \pm 0.3 \% \big) .
\eeqa
 
So far, our numerical estimates were based on the theoretical result for 
the form factor up to $\cO(p^4)$ in the strong part and to $\cO(e^2 p^2)$ 
for the electromagnetic contributions. 
In the case of the pion scattering lengths, it has been shown explicitly 
\cite{KU98} that the strong interaction contributions of $\cO(p^6)$  are of 
comparable size to the electromagnetic corrections. (See also \cite{MMS97} 
for a discussion of electromagnetic effects in neutral pion scattering.)
A similar feature can also be expected for the $K_{\ell 3}$ decays. 
Although a calculation of isospin conserving corrections of $\cO(p^6)$ is 
under way \cite{bi01}, for the time being we have to be content with the 
rough estimate \cite{lr84}
\beq \label{fplusp6}
f_+^{K^+ \pi^0} \big|_{p^6} = -0.016 \pm 0.008 
\eeq
given by Leutwyler and Roos already some years ago.
Adding\footnote{Also the input parameters in the strong $\cO(p^4)$ part 
of the form factor will receive appropriate shifts \cite{ABT2000} in a 
complete and 
consistent analysis of $\cO(p^6)$ effects.} (\ref{fplusnum}) and 
(\ref{fplusp6}), we 
obtain \beq \label{f+num}
f_{+}^{K^+ \pi^0} (0) = 0.9874 \pm 0.0084 , 
\eeq
and, consequently,
\beqa
|V_{us}| &=& 0.2207 \pm 0.0013 \pm 0.0008 \pm 0.0019 \nn
         &=& 0.2207 \pm 0.0024 ,
\eeqa 
which is in agreement with the current PDG value \cite{PDG},
\beq
|V_{us}| = 0.2196 \pm 0.0023 ,
\eeq 
taken from the analysis in \cite{lr84}. 

On the other hand, the unitarity of the Kobayashi--Maskawa 
mixing-matrix together with the present experimental value \cite{PDG} for 
$|V_{ud}|$ implies
\beq
|V_{us}| = 0.2287 \pm 0.0034 ,
\eeq 
indicating a possible problem for three-generation mixing.  Whether
this discrepancy will survive once two-loop chiral corrections are
included remains to be seen. It is interesting to notice that although
these corrections would have to be about twice the size of the
estimate (7.26) in order to maintain the unitarity of the CKM matrix,
they still would not be unnaturally large from the point of view of
the chiral expansion, and could be accounted for by ${\cal O}(m_s^2)$
contributions \cite{fks00}.

\section{Conclusions}
\label{sec: Conclusions}
\renewcommand{\theequation}{\arabic{section}.\arabic{equation}}
\setcounter{equation}{0}

In the present work, we have presented general formulae for the
$K_{\ell 3}$ form factors at one loop in the presence of radiative
corrections. Two features induced by the electromagnetic interactions 
are worth noticing. First, the form factors now depend not only on the
momentum transfered between the kaon and pion, but also on a second
kinematical variable. Second, in addition to the usual electromagnetic
low energy constants ($K_i$'s of Ref.~\cite{urech}), there are
contributions from four of the local counterterms $X_i$ introduced in
\cite{lept}.  These counterterms are specific to semileptonic
processes of pseudoscalar mesons and renormalize the ultraviolet
divergences induced by the exchange of virtual photons between charged
mesons and leptons.

Loops with virtual photons generate also infrared divergences. In
order to deal with them, we have analysed the associated real photon
emission processes.  We have given a general description of the
changes induced in the Dalitz plot density, and have proposed a
model-independent procedure for including radiative corrections in the
data analysis.  This consists in incorporating the known long-distance
electromagnetic effects into generalized kinematical densities, while 
including all the structure dependent (UV sensitive) terms as
corrections to the form factors.  Details of the new kinematical
densities will depend eventually on the way the specific experimental
set-up deals with real photon emission.  For a quite generic set-up
configuration (see Sect.~\ref{sec: corr}), we have rederived explicit
expressions for $K^+_{e3}$ mode, correcting some mistakes in earlier
work \cite{gin70}.

Within this framework, we have studied the effect of radiative
corrections in the extraction of the CKM matrix element $\vert
V_{us}\vert$ from the $K^+_{e3}$ mode.  As compared to the pure ${\cal
O}(p^4)$ form factors \cite{gl852}, the inclusion of ${\cal
O}(e^2p^2)$ electromagnetic contributions shifts $f_+(0)$ by about 
$(0.36 \pm 0.16)$\%.
Moreover, the radiative corrections produce an
effective reduction of -1.27\% for the phase space integral.  We note
here that the uncertainty on the form factor up to ${\cal
O}(p^4,e^2p^2)$ is well under control, and it affects the extraction
of $\vert V_{us}\vert$ only marginally compared to present
experimental errors. This feature persists in the case of $K^0_{e3}$
mode (not analysed numerically in this work), where the hadronic
uncertainties are even less effective. We think this is a   
relevant result of our analysis, opening the road to a precision  
determination of $\vert V_{us}\vert$, for which the next 
two important ingredients are: 
\begin{itemize}
\item From the theoretical side: the inclusion of two-loop
chiral corrections. 
\item From the experimental side: a new high statistics measurement of
branching ratios and slope parameters, including radiative corrections
in the model-independent way outlined in this work.
\end{itemize}

\medskip

\noindent
{\small {\it Acknowledgements.} We would like to thank G. Ecker for 
fruitful discussions and a careful reading of the manuscript. 
We are grateful to H. Pichl for useful comments.}

\appendix

\section{Mesonic Loop Functions}
\label{appA}
\renewcommand{\theequation}{\Alph{section}.\arabic{equation}}
\setcounter{equation}{0}

The loop function $H_{PQ} (t)$ \cite{gl852,GL85} is given by 
\begin{equation} 
H_{PQ} (t) =\displaystyle\frac{1}{F^2} \bigg[ h_{PQ}^r (t,\mu) 
+ \frac{2}{3}t L_9^r(\mu) \bigg] ~, 
\end{equation} 
where
\begin{eqnarray} 
h_{PQ}^r (t,\mu) &=& 
\frac{1}{12 t} \lambda 
(t,M_P^2,M_Q^2) \, 
\bar{J}_{PQ} (t) \nn
&&{}+ \frac{1}{18 (4 \pi)^2} (t - 3 \Sigma_{PQ}) \nn
&&{}- \frac{1}{12} \bigg\{ \frac{2 \Sigma_{PQ} - t}{\Delta_{PQ}} [A_P(\mu) 
- A_Q(\mu)] \nn
&& \qquad {}- 2 [A_P(\mu) + A_Q(\mu)] \bigg\}  , 
\end{eqnarray} 
with
\beq
\lambda (x,y,z)  =   x^2 + y^2 + z^2  - 2 ( x y + x z + y z )  , 
\eeq
\beq 
\Sigma_{PQ}  =  M_P^2 + M_Q^2~, \qquad \Delta_{PQ}  =  M_P^2 -
M_Q^2   , 
\eeq
\beq
A_P(\mu)   =   - \frac{M_P^2}{(4 \pi)^2} 
\log \frac{M_P^2}{\mu^2}   , 
\eeq
and
\beqa
\lefteqn{\bar{J}_{PQ} (t)  = 
\frac{1}{32 \pi^2} \Bigg[ 2 + 
\frac{\Delta_{PQ}}{t} 
\log \frac{M_Q^2}{M_P^2} - \frac{\Sigma_{PQ}}{\Delta_{PQ}} 
\log \frac{M_Q^2}{M_P^2} }  \nn
&&{}  - \frac{\lambda^{1/2} (t,M_P^2,M_Q^2)}{t} \nn
&& \times
\log  \left( \frac{[t + \lambda^{1/2} (t,M_P^2,M_Q^2)]^2 - 
\Delta_{PQ}^2}{[t - 
\lambda^{1/2} (t,M_P^2,M_Q^2)]^2 - \Delta_{PQ}^2} \right) \Bigg]   .
\eeqa
Moreover, in the expansion of the form factors  $\wt{f}_{-} (t)$, 
one needs the function: 
\beq
K_{PQ} (t) \,  = \,  \frac{\Delta_{PQ}}{2 t} \, \bar{J}_{PQ} (t)  \ . 
\eeq

\section{Photonic Loop Functions}
\label{appB}
\renewcommand{\theequation}{\Alph{section}.\arabic{equation}}
\setcounter{equation}{0}

The photonic loop contributions to the $K_{\ell 3}$ 
form factors depend on the charged lepton and meson masses 
$m_\ell^2$, $M^2$, as well as on the Mandelstam variables 
$u = (p_K - p_\ell)^2$ (for $K^{+}_{\ell 3}$ decays) and  
$s = (p_\pi + p_\ell)^2$ (for $K^{0}_{\ell 3}$ decays).
In what follows we denote by $v$ the Mandelstam variable 
appropriate to each decay.  
In order to express the loop functions in a compact way, 
it is useful to define the following intermediate variables:
\begin{equation}
 R  =  \frac{m_{\ell}^2}{M^2}  , \ \ 
 Y = 1 + R - \frac{v}{M^2}  , \ \     
 X = \frac{Y - \sqrt{Y^2 - 4 R}}{2 \sqrt{R}} . 
\end{equation} 
In terms of such variables and of the dilogarithm 
\begin{equation}
Li_2 (x) = - \int_{0}^{1} \frac{dt}{t} \log (1 - x t)   , 
\end{equation}
the functions contributing to $\Gamma_c (v,m_{\ell}^2,M^2 ; M_\gamma)$
and \\
$f_{\pm}^{\mbox{\small{EM-loop}}}(v)$ are given by
\beqa
\lefteqn{{\cal C} (v,m_\ell^2,M^2)  =   \frac{1}{m_\ell M} \frac{X}{1 - 
X^2}} \nonumber \\*
&\times&
\left[  - \frac{1}{2} \log^2 X + 2 \log X \log (1 - X^2) - 
\frac{\pi^2}{6} + \frac{1}{8} \log^2 R \right. \nn
&&{}+ \left.  Li_2 (X^2) + Li_2  (1 - \frac{X}{\sqrt{R}}) + 
Li_2 (1 - X \sqrt{R})  \right]  ,  \nonumber \\*
&&
\eeqa 
\beqa
\Gamma_1(v,m_{\ell}^2,M^2) &=& \frac{1}{2} \Big[ -\,\ln R\,+\,
(4-3Y){\cal F}(v,m_{\ell}^2,M^2) \Big]
\nonumber\\
\Gamma_2(v,m_{\ell}^2,M^2) &=& \frac{1}{2} \Big(1-\frac{m_{\ell}^2}{u}\Big) 
\Big[ - {\cal F}(v,m_{\ell}^2,M^2)(1-R)
\nonumber\\
&+& \ln R \Big]  -
\frac{1}{2}(3-Y){\cal F}(v,m_{\ell}^2,M^2)
\,,
\eeqa
and
\beq
{\cal F}(v,m_{\ell}^2,M^2)\,=\,\frac{2}{\sqrt{R}}\,\frac{X}{1-X^2}\,\ln X
\,.
\eeq

\section{Coefficients entering $\wt{f}_{-}^{K^+ \pi^0}$ and $\wt{f}_{-}^{K^0 
\pi^-}$}
\label{appC}
\renewcommand{\theequation}{\Alph{section}.\arabic{equation}}
\setcounter{equation}{0}

In this section we report the coefficients $a_{PQ}(t)$,
$b_{PQ}(t)$,\\
$c_{PQ}(t)$, and $d_{PQ}(t)$, appearing in the expressionf
of $\wt{f}_{-}^{K^+ \pi^0}$ and $\wt{f}_{-}^{K^0 \pi^-}$
(Eqs. (\ref{fminustildep}) and (\ref{fminustildez}) ).

\beqa 
a_{K^+ \pi^0} (t) & = & \frac{2 M_K^2 + 2 M_\pi^2 - t}{4 F_0^2} \nn  
&&{}+ \left(\frac{\ve^{(2)}}{\sqrt{3}}\right) \frac{-2 M_K^2 + 22 
M_\pi^2 
- 9 t}{4 F_0^2} 
+ 4 \pi \alpha Z  , \nn
a_{K^0 \pi^-} (t) & = & \frac{- 2 M_K^2 - 2 M_\pi^2 + 3 t}{2 F_0^2} 
\nn 
&&{}+ \left(\frac{\ve^{(2)}}{\sqrt{3}} \right) \frac{-2 M_K^2 + 6 
M_\pi^2 - 3 t}{2 F_0^2} 
- 16 \pi \alpha Z  , \nn
a_{K^+ \eta} (t) & = & \frac{2 M_K^2 + 2 M_\pi^2 - 3 t}{4 F_0^2} \nn 
&&{}+ \left(\frac{\ve^{(2)}}{\sqrt{3}}\right) 
\frac{6  M_K^2 - 2M_\pi^2 - 3 t}{4 F_0^2} + 12 \pi \alpha Z  .  \nn
&& 
\eeqa

\beqa 
b_{K^+ \pi^0} (t) & = & - \frac{M_K^2 - M_\pi^2}{2 F_0^2} - 
\left(\frac{7 \ve^{(2)}}{2 \sqrt{3}}\right) 
\frac{ M_K^2 - M_\pi^2}{F_0^2} 
- 4 \pi \alpha Z  , \nn
b_{K^0 \pi^-} (t) & = & - \frac{M_K^2 - M_\pi^2}{F_0^2} - 
\left(\frac{\ve^{(2)}}{ \sqrt{3}} \right) \frac{ M_K^2 - 
M_\pi^2}{F_0^2} 
- 8 \pi \alpha Z  , \nn
b_{K^+ \eta} (t) & = & - \frac{3}{2} \frac{M_K^2 - M_\pi^2}{F_0^2} +
\left(\frac{\sqrt{3} \,  \ve^{(2)}}{2}\right) \frac{ M_K^2 - 
M_\pi^2}{F_0^2}
\nn 
&&{}- 12 \pi \alpha Z  . 
\eeqa

\beqa 
c_{K^+ \pi^0} (t) & = & - \frac{2 M_K^2 + 2 M_\pi^2 - 3 t}{4 F_0^2} + 
\left(\frac{\ve^{(2)}}{\sqrt{3}} \right) 
\frac{- 4 M_K^2 + 3 t}{2 
F_0^2} \nn
&&{} - 8 \pi \alpha Z  , \nn
c_{K^0 \pi^-} (t) & = & \frac{t}{2 F_0^2}  , \nn
c_{K^+ \eta} (t) & = &  \frac{2 M_K^2 + 2 M_\pi^2 - 3 t}{4 F_0^2} 
+ \left( \frac{\ve^{(2)}}{\sqrt{3}} \right)
 \frac{ 4 M_K^2 - 3 t}{2 F_0^2}  . \nonumber \\*
&&
\eeqa

\beqa 
d_{K^+ \pi^0} (t) & = & - \frac{M_K^2 - M_\pi^2}{2 F_0^2} - 
\left(\frac{4 \, \ve^{(2)}}{\sqrt{3}}\right)  
\frac{ M_K^2 - M_\pi^2}{F_0^2} 
+ 4 \pi \alpha Z  , \nn
d_{K^0 \pi^-} (t) & = & - \frac{M_K^2 - M_\pi^2}{F_0^2} - 
\left(\frac{2 \, \ve^{(2)}}{ \sqrt{3}}\right) 
\frac{ M_K^2 - M_\pi^2}{F_0^2} 
+ 8 \pi \alpha Z  , \nn
d_{K^+ \eta} (t) & = & - \frac{3}{2} \frac{M_K^2 - M_\pi^2}{F_0^2}  
+ 12 \pi \alpha Z  . 
\eeqa

\end{document}